\documentclass[a4paper,11pt]{article}
\pdfoutput=1 

\usepackage{jinstpub} 
\usepackage{caption}
\usepackage{subcaption}
\usepackage{float}
\usepackage{makecell}
\usepackage[T1]{fontenc}
\usepackage[utf8]{inputenc}
\usepackage{upgreek}
\usepackage{csquotes}
\usepackage{natbib}

\title{Characterization of a beam-tagging hodoscope for hadrontherapy monitoring}

\author[a,1]{O. Allegrini,\note{Corresponding author.}}
\author[b]{J.-P. Cachemiche,}
\author[b]{C.P.C. Caplan,}
\author[a]{B. Carlus,}
\author[a]{X. Chen,}
\author[c]{S. Curtoni,}
\author[c]{D. Dauvergne,}
\author[a]{R. Della Negra,}
\author[c]{M.-L. Gallin-Martel,}
\author[e]{J. H\'{e}rault,}
\author[d]{J.M. L\'{e}tang,}
\author[b]{C. Morel,}
\author[a]{\'{E}. Testa,}
\author[a]{and Y. Zoccarato.}

\affiliation[a]{Univ. Lyon, Univ. Claude Bernard Lyon 1, CNRS/IN2P3, IP2I Lyon, F-69622, Villeurbanne, France.}
\affiliation[b]{Aix-Marseille Univ, CNRS/IN2P3, CPPM, Marseille, France.}
\affiliation[c]{Universit\'e Grenoble Alpes, CNRS, Grenoble INP, LPSC-IN2P3, UMR 5821, 38000 Grenoble, France.}
\affiliation[d]{Univ. Lyon, INSA-Lyon, Univ. Claude Bernard Lyon 1, UJM-Saint \'Etienne, CNRS, Inserm, CREATIS UMR 5220, U1206, F-69373, LYON, France.}
\affiliation[e]{Department of Radiation Oncology, Antoine-Lacassagne Cancer Center, Nice, France.}

\emailAdd{o.allegrini@ipnl.in2p3.fr}

\abstract{A beam tagging hodoscope prototype made of squared 1~mm$^2$ fibers arranged in two perpendicular planes and coupled to multi-anode photomultipliers has been studied using 65 MeV proton as well as 95 MeV/u $^{12}$C beams at various intensities. This hodoscope successfully provided 2D images of proton beams with a detection efficiency larger than 98\% with logical OR condition between the two fiber  planes. The detection efficiency with a coincidence between the two planes is close to 75\% for beam intensities up to $\sim1$~MHz. Moreover, the timing resolution is around 1.8~ns FWHM. Overall, the performances show that such a technology is viable for beam monitoring during hadrontherapy. }


\usepackage{xcolor}

\begin{document}
\maketitle
\flushbottom

\section{Introduction}
\label{sec:intro}

Ion beam therapy is a rapidly expanding radiotherapy modality, with more than 250,000 patients being treated worldwide\footnote{\url{https://www.ptcog.ch/index.php/patient-statistics}}. The main advantage of this technique lies in the dose conformity with the target volume resulting from the sharp Bragg peak observed in the dose profile at the end of the ion range. Moreover, therapies with ions heavier than protons benefit from increased biological effect in the tumor region, hence enhancing the treatment effectiveness \cite{Braccini2010, Durante2016, Schardt2010, Paganetti2013, Jakel2008}. However, this therapy technique faces uncertainties concerning the Bragg peak position mainly due to X-ray imaging modalities, anatomical changes of the patient during the treatment, organ motion and approximations used in dose calculation \cite{Paganetti2012}. 
As a consequence, the most widespread treatment planning techniques are performed with several beam incidences accommodating treatment robustness at the expense of higher doses in healthy tissues. Moreover, additional safety margins are applied around the tumor volume to ensure the full irradiation of  tumor cells \cite{Durante2016, Knopf2013}.

In this context, clinically applicable methods and instruments are under development to monitor \textit{in vivo} ion ranges and dose profiles with millimeter accuracy using secondary radiation detection. 
Some detection systems exploit the production of $\beta+$ emitters during nuclear reactions undergone by a fraction of incident ions. At present, ion-range verification is only performed after treatment in some hadrontherapy centers employing commercial PET/CT scanners. But in-beam PET scanners are currently developed and tested in clinical conditions to provide online ion-range monitoring \cite{Shao2014, Ferrero2018}.
Besides the production of $\beta+$ emitters, nuclear reactions also lead to the emission of prompt gammas (PG) that can be also considered for ion-range verification \cite{Krimmer2018}. Several PG modalities are under development or in their clinical setting evaluation worldwide and a few of them make use of time-of-flight (TOF) measurement either to derive indirect information on ion-ranges (Prompt Gamma Timing, PGT \cite{Golnik2014, Marcatili2020}) or to reduce neutron-induced background in PG imaging systems (collimated and Compton cameras) \cite{Fontana2020, Dal_Bello_2020, Aldawood2017}. 

TOF measurement requires a time reference that cannot be provided by beam monitoring systems implemented in clinical facilities (based on ionization chambers \cite{Stelzer2002}) since they are not designed to provide such a timestamp. The solution consisting in using the accelerator radio-frequency (RF) as time reference would have the advantage of simplicity, when there are a perfect periodicity and short ion bunches (cyclotron accelerators). However, the precise correlation between the RF phase and the ion/bunch arrival can be obtained in mono-energetic beam conditions only. Indeed the use of degraders to change beam energies in cyclotrons introduces time dispersion and shifts of bunches. Moreover, small variations of the cyclotron’s magnetic field slightly affect the orbital frequency of the ion trajectories. Hence a small phase shift can occur at each turn between the ion trajectory and the RF signal, which results in a time-varying measurable mismatch \cite{Petzoldt2016}. Several alternative solutions are currently being studied to provide a more accurate time reference corresponding to the arrival of incident ions or ion bunches.

Among them, some devices under development based on scintillating fibers provide particles tracking with integration \cite{Leverington2018} or particle-per-particle \cite{Horikawa2004, Achenbach2008, Braccini2012} acquisition mode for various fields of application, including proton radiography \cite{Presti2016} and ion-range verification during hadrontherapy \cite{PAPA2016}. Recently, the performance of a time-tracker for a prompt-gamma spectroscopy system allowing for a background TOF rejection with a sub-nanosecond time resolution has been demonstrated \cite{Martins2020}. Among the scintillating fibers hodoscope currently developed, the one of the ClaRyS collaboration is dedicated to be coupled with a PG imaging system (collimated or Compton camera). The desired detection efficiency of the beam tagging hodoscope should be around 90\% for coincidence events in the X and Y planes to ensure a spatio-temporal tagging of a maximum number of incident ions in order to minimize the loss of statistics. The time resolution must remain below 2~ns FWHM, which corresponds to the bunch width of a clinical cyclotron. Furthermore, the device should be compatible with clinical conditions both in terms of beam intensities leading to counting rates up to 100~MHz, which is almost the frequency at which the ion bunches are delivered in a cyclotron facility, and radiation hardness (device operational for at least 1000 patient treatments, considering an average dose of 60 Gy per treatment).

The present paper reports on in-beam tests of the hodoscope developed by the CLaRyS collaboration. At first, performance tests were performed at GANIL with 95~MeV/u carbon ions to measure the detection efficiency as well as the variation of the response of the scintillating fibers as a function of the beam fluence. The second part of the paper presents the setting method of the data acquisition system (thresholds and gains) and the results of the in-beam characterization performed during two experiments at the Mediterranean Protontherapy Institute, in Nice. Multiplicity, detection efficiency and time resolution were assessed. 

\section{Material and methods}
\subsection{The beam tagging hodoscope}
\label{sec:hodo}

The beam tagging hodoscope of the CLaRyS collaboration is designed to provide a spatio-temporal tagging of ion (synchrotrons) or ion bunches (cyclotrons) passages for counting rates up to 100~MHz, which corresponds to a period of 10~ns, the typical period of cyclotron beam microstructures in hadrontherapy centers. 

\begin{figure}[htb]
\centering
\begin{subfigure}{0.4\textwidth} \centering
\includegraphics[width=\textwidth]{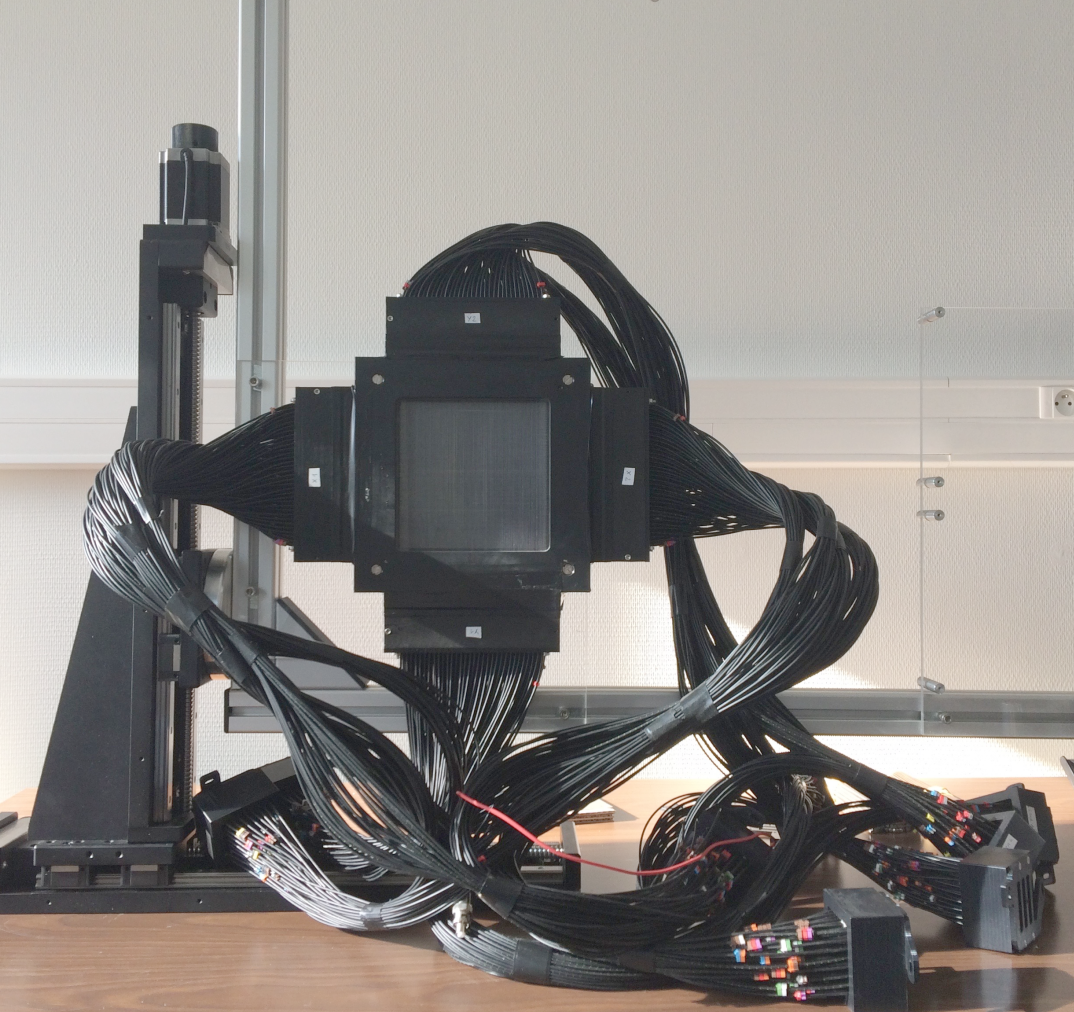}\caption{}
\label{fig:hodoscope}
\end{subfigure}
\begin{subfigure}{0.4\textwidth} \centering
\includegraphics[width=\textwidth]{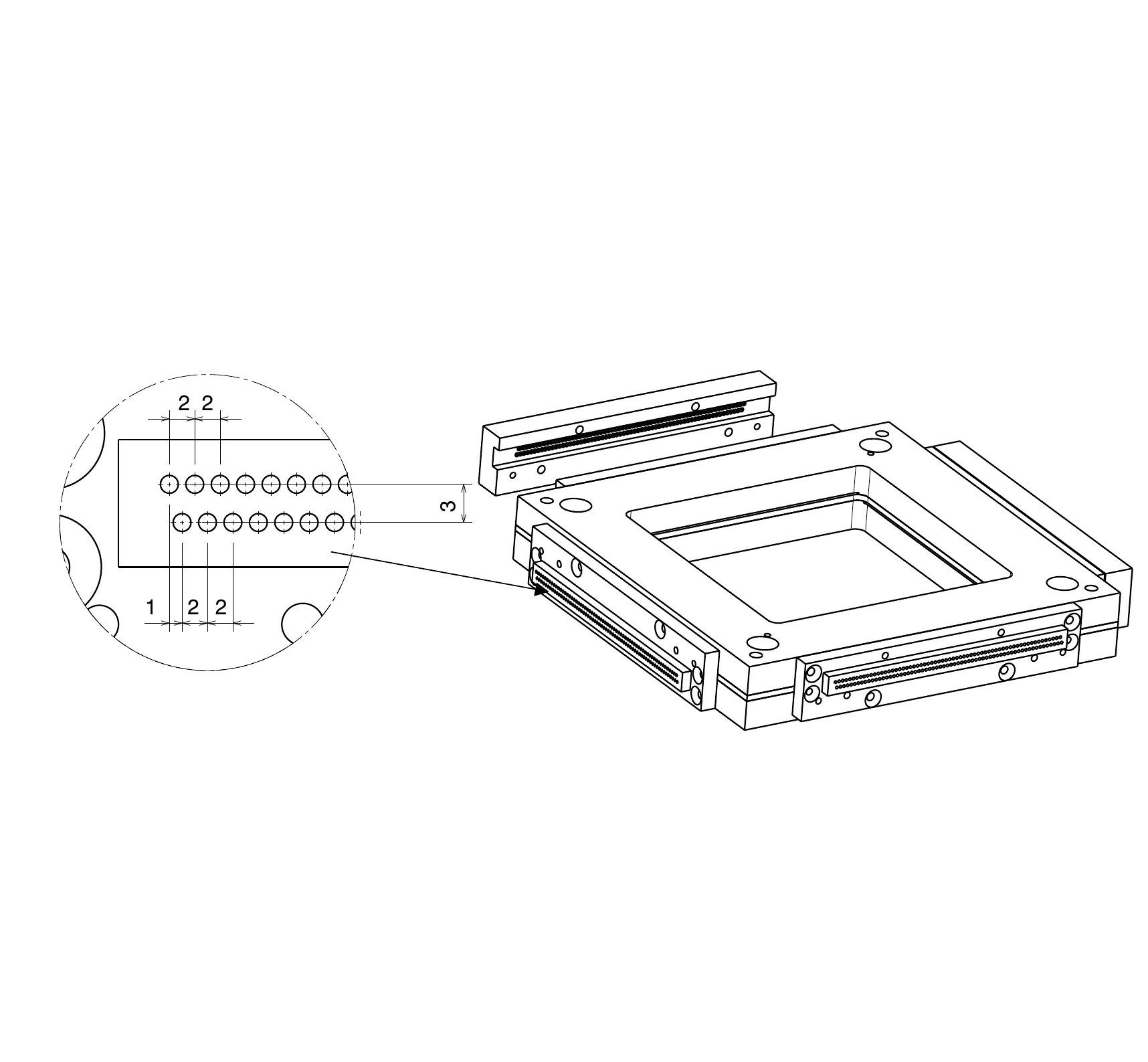}\caption{}
\label{fig:frame_hodoscope}
\end{subfigure}
\caption{\small{\textit{(a) Large version of the beam tagging hodoscope. (b) Scheme of the frame of the hodoscope and eyelets arrangement.}}}
\label{fig:Hodoscope}
\end{figure}

The final version of the beam tagging hodoscope is composed  of 1~mm$^{2}$ square-section polystyrene scintillating fibers BCF-12 manufactured by Saint-Gobain \cite{SaintGobain2017}.
The beam tagging hodoscope is constituted of two perpendicular planes of fibers. Each plane contains 128 fibers, which gives an active area of 128$\times$128~mm$^{2}$ (Figure~\ref{fig:hodoscope}). The scintillating fibers are kept together with a transparent glue and accurate positioning of the fibers in the layers is ensured by the frame of the hodoscope (Figure~\ref{fig:frame_hodoscope}) and the eyelets pierced through it where they are coupled with optical fibers. During operation, the sensitive part is isolated against external light. Fibers are readout on both sides by 8 Hamamatsu multi-anode photomultiplier tubes (PMTs) H8500C. The number of read-out channels is then 512. Each PMT is segmented in 64 pixels of 5.8$\times$5.8 mm$^2$ and every fiber sides are fixed to the PMTs anode surfaces through a plastic custom mask. Connections with fibers are made in such a way that every fiber is read-out by two neighboring channels of a PMT and two adjacent fibers are handled with different PMTs. This configuration allows for optimizing detector efficiency and time resolution. 

\begin{figure}[htb]
\centering
\includegraphics[width=\textwidth]{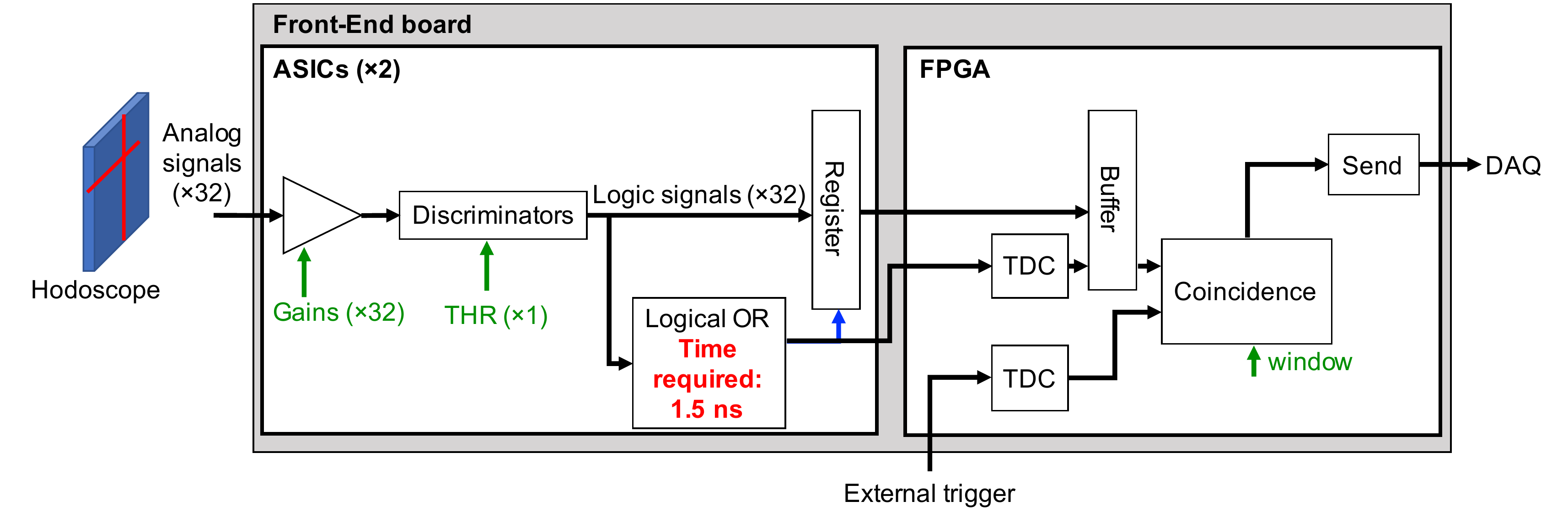}
\caption{\small{\textit{Simplified flowchart of the signal fibers processing in the FE card. The flow of signals and data is represented with black arrows while parameters are in green and writing order in blue.}}}
\label{fig:Scheme_reading_hodo}
\end{figure}

Each PMT is linked to a front-end (FE) card via a 64-channel connector. The main components of this card are two 32-channel readout ASICs \enquote{HODOPIC}, a signal-processing FPGA, a single-channel optical transceiver and an RJ45 connector \cite{Chen2019}. As detailed in Figure~\ref{fig:Scheme_reading_hodo}, the ASICs provide logic signals associated to each channel and the logical OR of these signals, that triggers the storage of the channel states in an ASIC register. A specific gain is assigned to each channel while a common threshold is applied on all channels of one ASIC. When a logical OR is generated, the duration of the logic signals corresponds to the time for which the signal amplitude is larger than the threshold. It is worth noting that it has to be larger than the time required to generate the logical OR which is about $1.5$~ns. Otherwise, when the state of the channels is not stored in the ASIC, the channels are considered to be not hit .

Apart from signal processing, the FPGA has to determine the time difference between the trigger signal sent by the gamma camera and the time stamp associated to the hodoscope. This time difference is indeed required for the TOF measurement. The time stamps associated to the two signals (trigger and hodoscope) is determined by three time-to-digital converters (TDC), one for the trigger and one for each ASIC. The time stamp value of the hodoscope corresponds to the lowest value measured by the two TDCs associated to the ASICs. Their time resolution (LSB, Least Significant Bit) is of 0.3125~ns (hodoscope) and 0.625~ns (trigger). A single time difference value is therefore provided per ASIC for which at least one channel has been hit.

\begin{figure}[htb]
\centering
\includegraphics[width=1\textwidth]{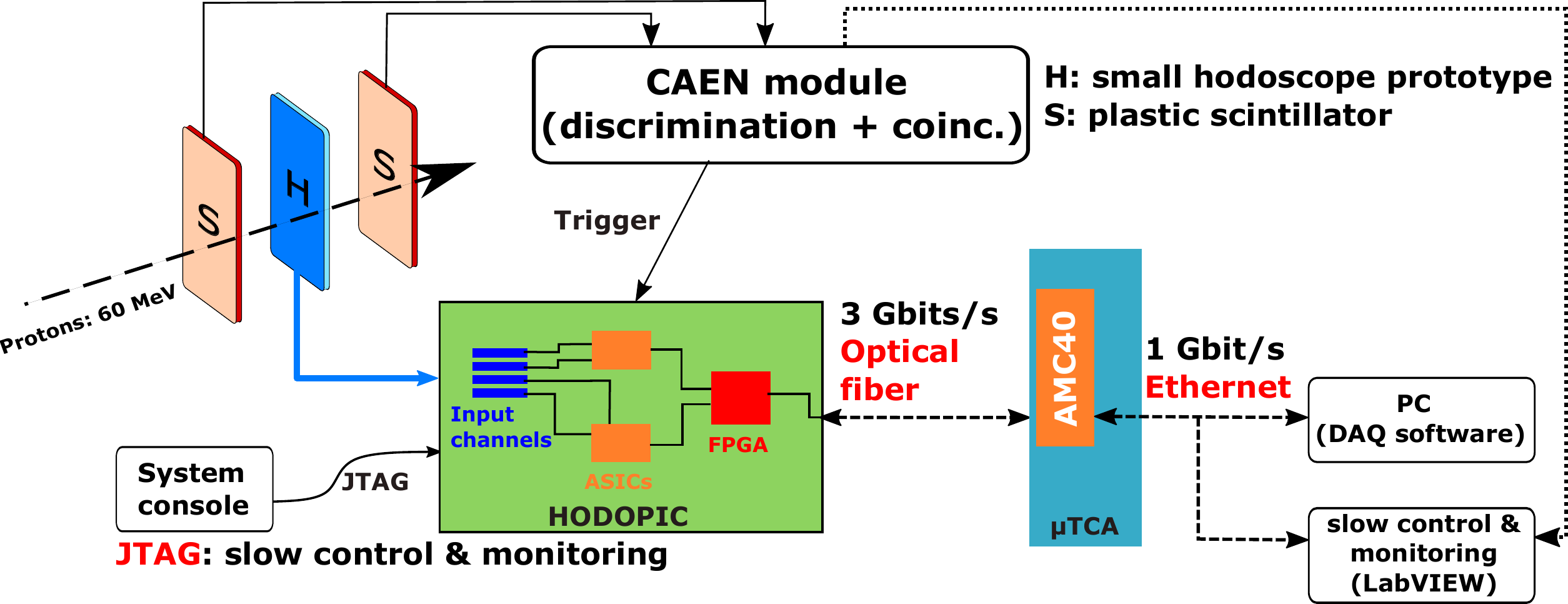}
\caption{\small{\textit{Scheme of the data acquisition chain of the experimental setup.}}}
\label{fig:Scheme_Setup_hodo}
\end{figure}

If it falls within a given time window, the data are sent from the FPGA to the back-end card (AMC40) \cite{Cachemiche2010} through a 3~Gbit$\cdot$s$^{-1}$ optical fiber with a specific protocol \cite{deng2013, Chen2017, Chen2019, Caplan2019} and then to the Personal Computer (PC) through 1~Gbit$\cdot$s$^{-1}$ ethernet link where data are processed and stored by the data acquisition software. In the case of the nominal acquisition mode, the data consists of the list of hit fibers with the fiber number and the associated time-of-arrival (TOA). Both optical and ethernet links also transmit slow-control packages sent by a LabVIEW-based program. Figure~\ref{fig:Scheme_Setup_hodo} shows schematically the experimental setup in which the whole acquisition chain is illustrated with front-end (HODOPIC) and back-end (AMC40) cards, the slow control system  and the data acquisition software.

In order to monitor the ageing of the fibers, another data acquisition mode has been foreseen in which the signal amplitude of a given channel is measured using an ADC implemented in the ASIC.

For this study, a smaller hodoscope with 32 fibers per plane has also been developed in order to use a single acquisition board to collect all the data of the two planes. For this small prototype, each fiber plane is readout from a single side and by a single ASIC. 

Overall, the data acquisition parameters of the hodoscope are the following: the PMT high-voltage and the ASIC gains and thresholds. As detailed in section~\ref{sec:Settings}, these parameters can be optimized to improve the data collection and the signal-to-background ratio.

\subsection{General principle of the experimental setups}
\label{GeneralPrinc}

\begin{figure}[htb]
\centering
\includegraphics[width=0.5\textwidth]{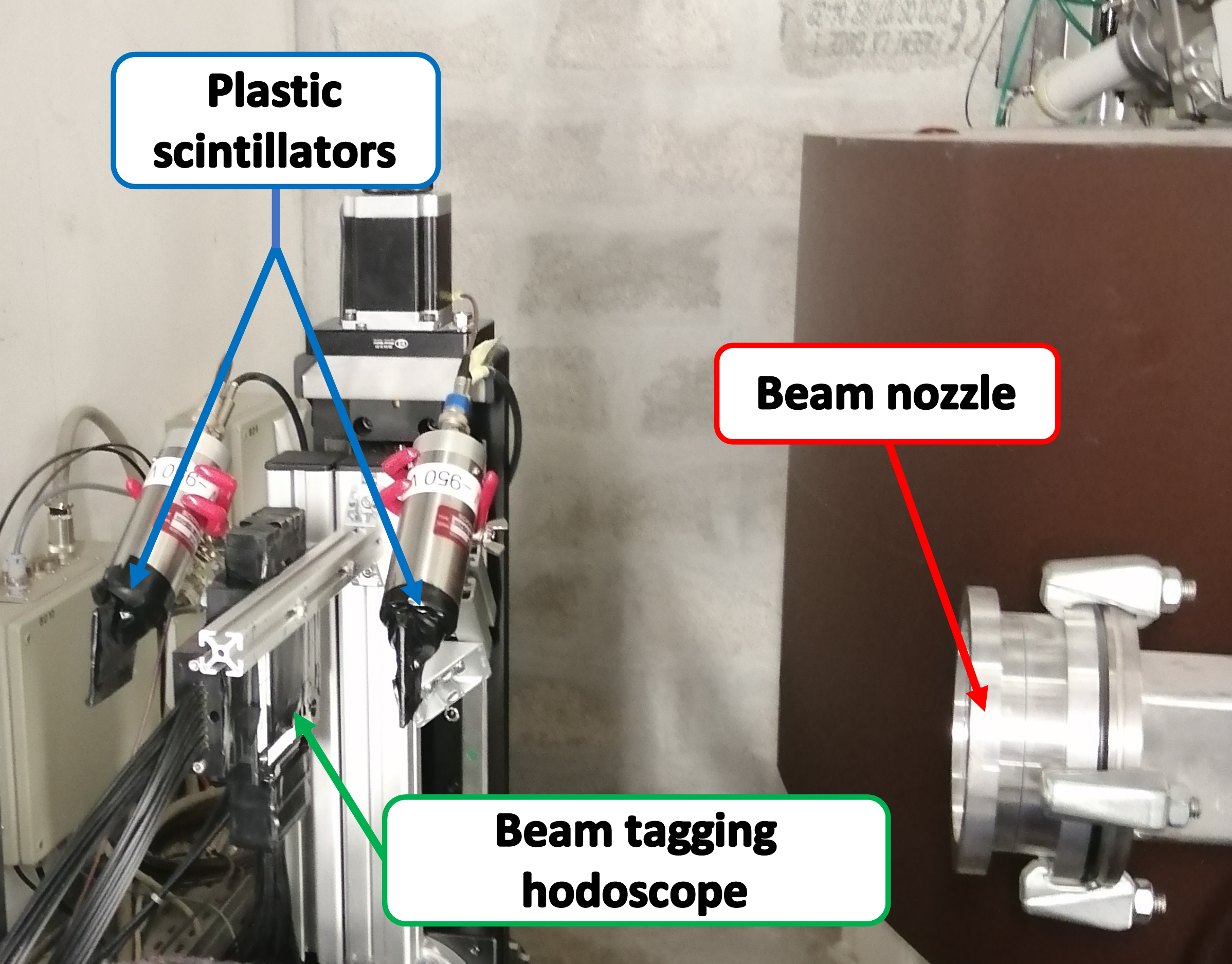}
\caption{\small{\textit{Typical experimental setup used to assess the performance of the hodoscope during in-beam tests. The two plastic scintillators are used in order to provide an external trigger signal when a proton impinges the hodoscope.}}}
\label{fig:Picture_Setup_hodo}
\end{figure}

In the final configuration of the CLaRyS camera, the beam tagging hodoscope is coupled to a gamma detector in which the PG are absorbed. A trigger signal is generated by the absorber upon the detection of a PG. This trigger signal is then sent to the FPGA of the hodoscope FE card. The characterization of the hodoscope efficiency requires a suitable device to detect the protons passing through the hodoscope. For this reason, in this experimental study, the trigger signal provided by the absorber of the CLaRyS prototype is replaced by an external trigger obtained by the coincidence signal of two plastic scintillators (PSs) surrounding the hodoscope. Figure \ref{fig:Picture_Setup_hodo} illustrates the setup used to assess the performances of the hodoscope during in-beam tests. 

The fact that these detectors are slightly larger than the beam hodoscope does not matter since we have verified that the whole beam crosses the hodoscope (Figure~\ref{fig:2D_Maps} in the section \ref{Results}). PSs are aligned with the beam and located about 5~cm upstream and downstream of the hodoscope. Their signals are read out by PMT which are connected to a Nuclear Instrumentation Module (NIM). This module ensures the analog to logic signal conversion thanks to a preset threshold. This configuration allowed us to generate double (two PSs) and triple coincidences (two PSs AND the beam tagging hodoscope). A logic signal is generated
by the hodoscope when one (logical OR) or two fiber planes (logical AND) detect a particle. The ratio of the number of triple and double coincidences is a direct measurement of the detector efficiency.

\subsection{GANIL experiments}

The setup described in the previous section was used to measure the hodoscope efficiency with  95~MeV/u carbon beams at low intensities (20--30~pA) and to assess radiation damage. The single Hamamatsu multi-anode PMT connected to the small hodoscope prototype was biased at 800~V. Its readout was performed with various NIM modules to provide trigger signals (discriminator module), charge integration (QDC module) and timing (TAC module)\footnote{QDC: charge-to-digital converter : TAC: time-to-amplitude converter.}. The discriminator threshold was set at 30~mV, which corresponds to about one sixth of the maximum of the amplitude distribution. The effect of radiation damage was estimated by comparing the detection efficiency measured for different fluences with high beam intensities (2--3~nA).

\subsection{Centre Antoine Lacassagne (CAL) experiments}
\label{In-beam_tests}

Since the counting rate capability of PS is around {10}$^{5}$~Hz, a beam current monitor (BCM) \cite{Kelleter2017, Martins2020} was used to monitor larger beam intensities. It consists of a scintillator placed out of the beam irradiation field and calibrated with the in-beam PS at low beam intensity and with the intensity measured in the cyclotron stripper at high intensities. 

Moreover, the three single independent signals and the coincidence module output signal were sent to the LabVIEW-based program to visualize the various counting rates and to provide online monitoring of the beam intensity. 
Finally, the time selection window applied in the FPGA of the FE card, before data transmission, was tuned from an additional PC through a JTAG\footnote{JTAG: Joint Test Action Group.} link.

The MEDICYC low energy treatment line of the CAL is intended for the treatment of ocular tumors. The research area of this beam line is located a few meters upstream of the treatment room. The maximum beam energy is 64.5~MeV and the high frequency of the accelerator is 24.85~MHz so that the beam consists of proton bunches arriving every 40.24~ns. Throughout the rest of the paper, beam intensities are expressed in Hz because the mean number of protons per bunch is always below 1 for the range of intensities (much below typical clinical ones) used during these experiments. The counting rates recorded in the PSs are approximated to units of Hz.
The performance of the hodoscope is mainly assessed in terms of detection efficiency and spatial resolution. The number of hit fibers per plane when a trigger is generated (multiplicity $M$) is also considered in this characterization study as well as in the detection profiles.

\subsection{Criteria and settings}
\label{sec:Settings}

As mentioned in section~\ref{sec:hodo}, the data acquisition parameters of the hodoscope are the PM high-voltage and the ASIC gains and thresholds. Their settings aim at finding the best compromise between noise rejection and detection efficiency.

The standard method consists in measuring the so-called S-curves obtained using a logic periodic signal sent to each input channel, one by one. The S-curve corresponds to the number of pulses detected by the FE card over a given acquisition time as a function of threshold. Actually, two types of events can be distinguished according to the various signals provided by the ASIC, namely the logic signals associated to each channel and the request signal (logical OR of all channels):
\begin{itemize}
  \item good events (GEs): events for which at least one logic signal state is high, 
  \item events with short pulse (ESPs): undesired events for which the logic signal is not long enough to have a high state when the reading of the channel states by the FPGA is requested by the ASIC (see section~\ref{sec:hodo}).
\end{itemize}

\begin{figure}[htb]
\centering
\includegraphics[width=0.5\textwidth]{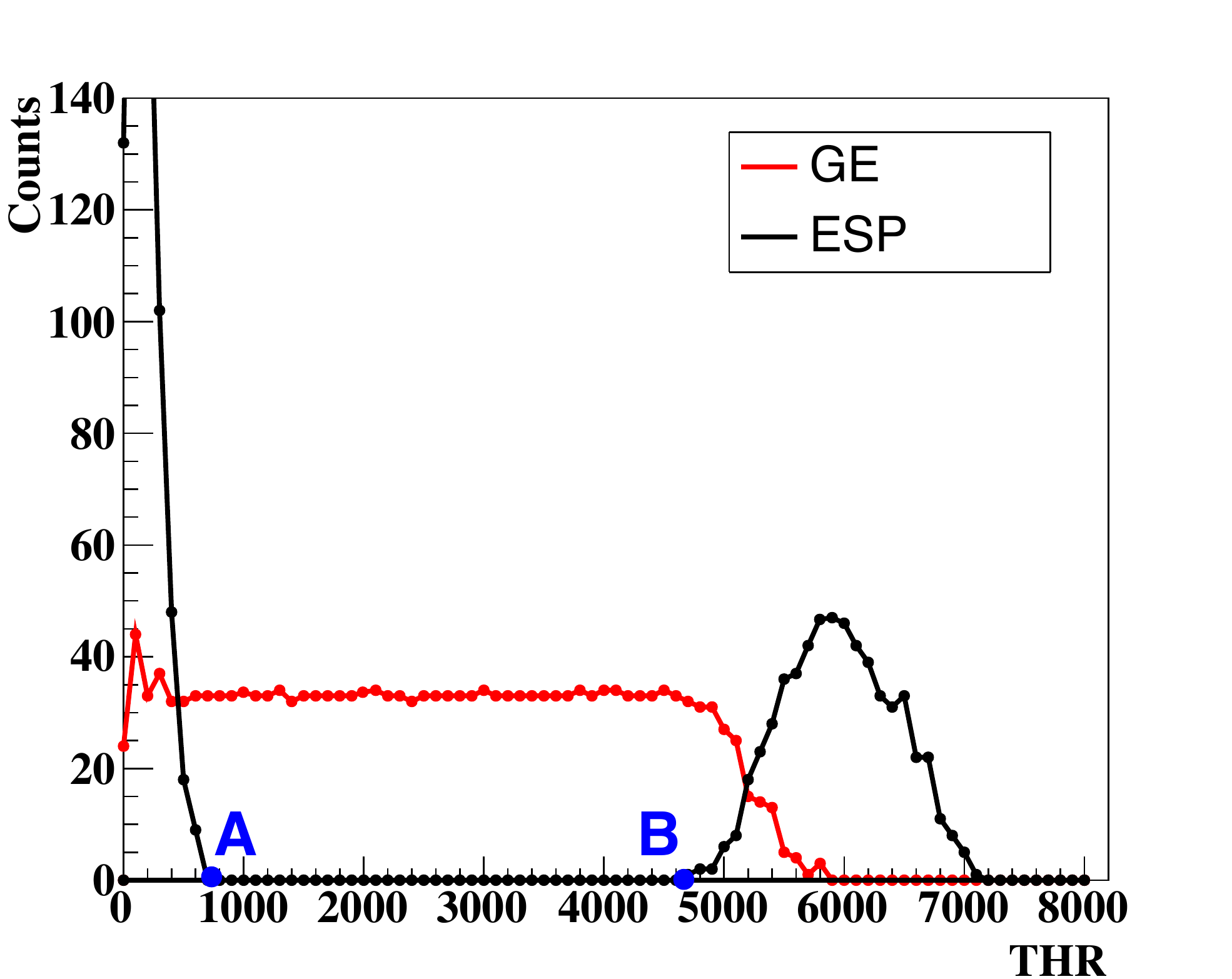}
\caption{\small{\textit{Number of GEs and ESPs events as a function of the threshold (THR) for a given channel at a gain value of 1.50. The range between points A and B corresponds to a suitable threshold range for this channel.}}}
\label{fig:S_Curve}
\end{figure}

For illustration purpose, a typical S-curves is shown in Figure \ref{fig:S_Curve}. As expected the GE curve presents a plateau (the expected number of pulses during the acquisition time) up to the sharp fall-off for threshold values close to the signal amplitude (point B). In this fall-off region, the larger the threshold the smaller the probability to have long enough logic signals to be readout by the FPGA. This is the reason why the ESP curve presents a maximum at the bottom of the GE curve fall-off. Finally the peak observed in the ESP curve for very low threshold values (below point A) is due to the electronic noise. The interval defined by points A and B corresponds to the suitable threshold range (STR) for a given channel gain.

\begin{figure}[htb]
\centering
\includegraphics[width=\textwidth]{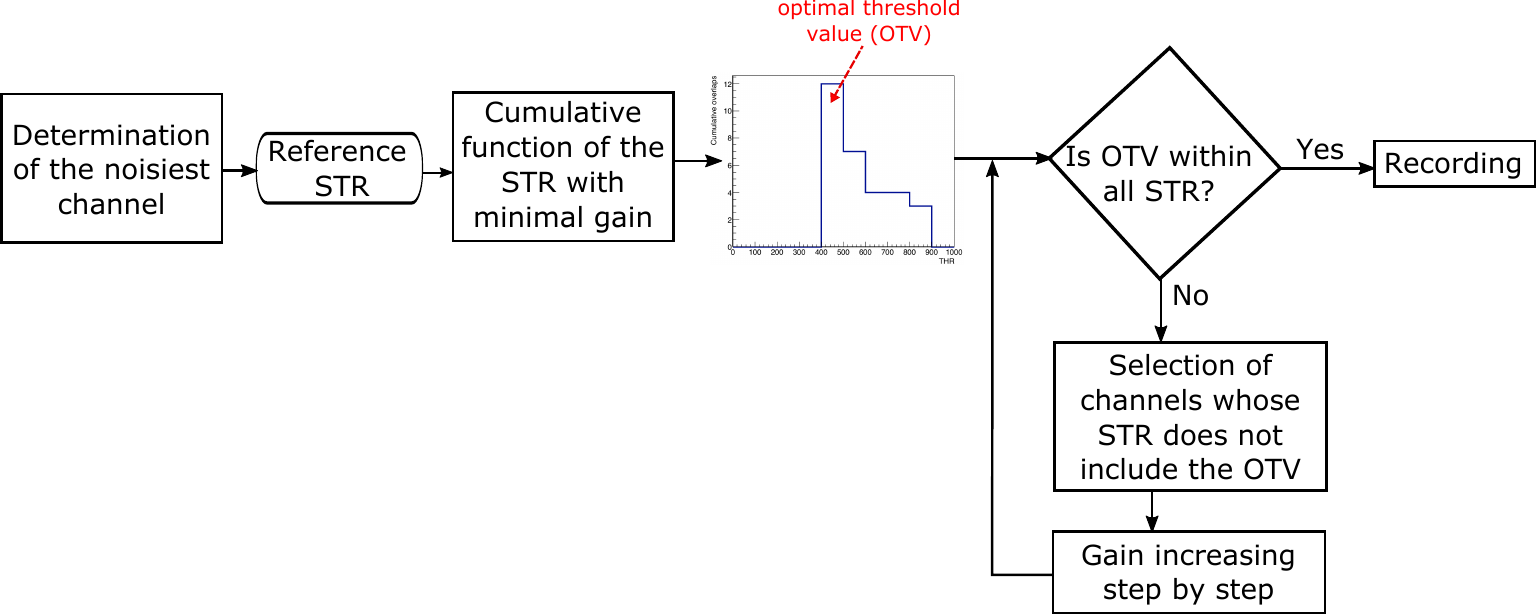}
\caption{\small{\textit{Block diagram of the method used to set the threshold value of the ASICs and the gain of the channels: the optimal threshold value (OTV) is obtained from the cumulative function of the suitable threshold range (STR). Gains are adjusted next to the OTV determination.}}}
\label{fig:Simplified_algo}
\end{figure}

The method used to determine a set of gains and threshold is described in Figure~\ref{fig:Simplified_algo}. It is based on three main steps:
\begin{itemize}
	\item a first scan of the input channels consists in determining the noisiest channel for the minimal gain (0.25), i.e.~the channel for which point A (in figure \ref{fig:S_Curve}) corresponds to the largest threshold value. Points A and B of this channel are then used as a reference STR;
	\item the STR of the other channels is then measured for the minimal gain (0.25). The cumulative function of the overlap of this STR with the reference STR is then computed. The maximum of this distribution is defined as the optimal threshold value;
	\item for channels whose STR does not include the optimal threshold value, the gain is increased step by step until the optimal threshold value shows up in the STR.
\end{itemize}

Finally, the cross-talk between neighboring pixels has been assessed by scanning the PMTs with a blue LED. This LED is mounted on a motorized double-axis table having a step resolution of 20~$\upmu$m. It produces light pulses synchronized with a pulse generator, which are split in two pulses with a mirror. One is sent to the H8500C PMT via an optical fiber in order to obtain a light spot perpendicular to the cathode surface and the other to an Hamamatsu R5600 PMT used as reference for the correction of temperature fluctuations. Although slightly larger than the provider specifications, the cross-talk measured with our setup always remains below 3\%. This fraction of cross-talk events can therefore be considered as relatively low. Moreover, since the associated signal amplitude is relatively low, one can expect that this source of background is cut with thresholds determined by the setting of the ASIC parameters (gains and threshold) \cite{FontanaPhD}.

\section{Results}
\label{Results}
\subsection{Experiment with carbon ions at GANIL}
A total detection efficiency of 94\% was obtained with a single plane, while the fraction of events with multiplicity $M>1$ represented 24\%, mainly due to cross talk between adjacent channels of the multi-anode PMT. From SRIM estimation \cite{Ziegler2010}, the energy deposit in a 1 mm thick plastic fiber is 26 MeV with a 95~MeV/u carbon ion beams. The results for the integrated charge per projectile, and the measured efficiency, are reported in Table~\ref{tab:GANIL}. At the beginning of the experiments, the detection efficiency of 94\% was measured after a few short low-intensity irradiations. A significant decrease of the efficiency was measured after 2.2$\times$10$^{12}$~ions$\cdot$cm$^{-2}$, which represents more than 1000 patient treatments, but it was almost recovered by lowering the threshold.
Close to 90\% detection efficiency was kept after 3.6$\times$10$^{12}$ ions$\cdot$cm$^{-2}$.
Concerning the irradiation damages, it should be noted that after few days, a partial restoration of the response at laboratory test benches has been observed, but it was not possible to repeat the detection efficiency measurement with carbon ions. 
The time resolution was measured between the logical OR of the Y plane and the cyclotron high-frequency signal and was assessed to 550~ps RMS. 
\begin{table}[htb]
\caption{\small{\textit{Evolution of the hodoscope response as a function of 95~MeV/u carbon ion fluence. DT: Discriminator threshold. The detection efficiency is measured on a single fiber plane with a multi-anode PMT H8500 operated at 800~V. A threshold of 30~mV corresponds to 4.5~pC measured on QDC. The first column with a null fluence corresponds to a negligible fluence after a few short low-intensity irradiations at the beginning of the experiments.}}}
\centering
\begin{tabular}{|c|c|c|c|c|c|}
\hline
Fluence (cm$^{-2}$)& $\sim$0 & (7.2$\pm$1.2)$\times$10$^{11}$ & (2.2$\pm$0.4)$\times$10$^{12}$ & (2.2$\pm$0.4)$\times$10$^{12}$ & (3.6$\pm$0.7)$\times$10$^{12}$\\
\hline
\makecell{DT (mV)} & \multicolumn{3}{c|}{30} &\multicolumn{2}{c|}{15}\\
\hline
\makecell{Mean QDC\\value (pC)} & 35 & 34 & 27 & 21 & 21\\
\hline
Efficiency (\%) & 94 & 94 & 63 & 92 & 86\\
\hline
\end{tabular}
\label{tab:GANIL}
\end{table}

\subsection{Experiments with protons at CAL}
\subsubsection{Profiles and multiplicities}
\label{Profiles_And_Multiplies}

\begin{figure}[htb]
\centering
    \begin{subfigure}{0.32\textwidth} \centering \includegraphics[width=\textwidth]{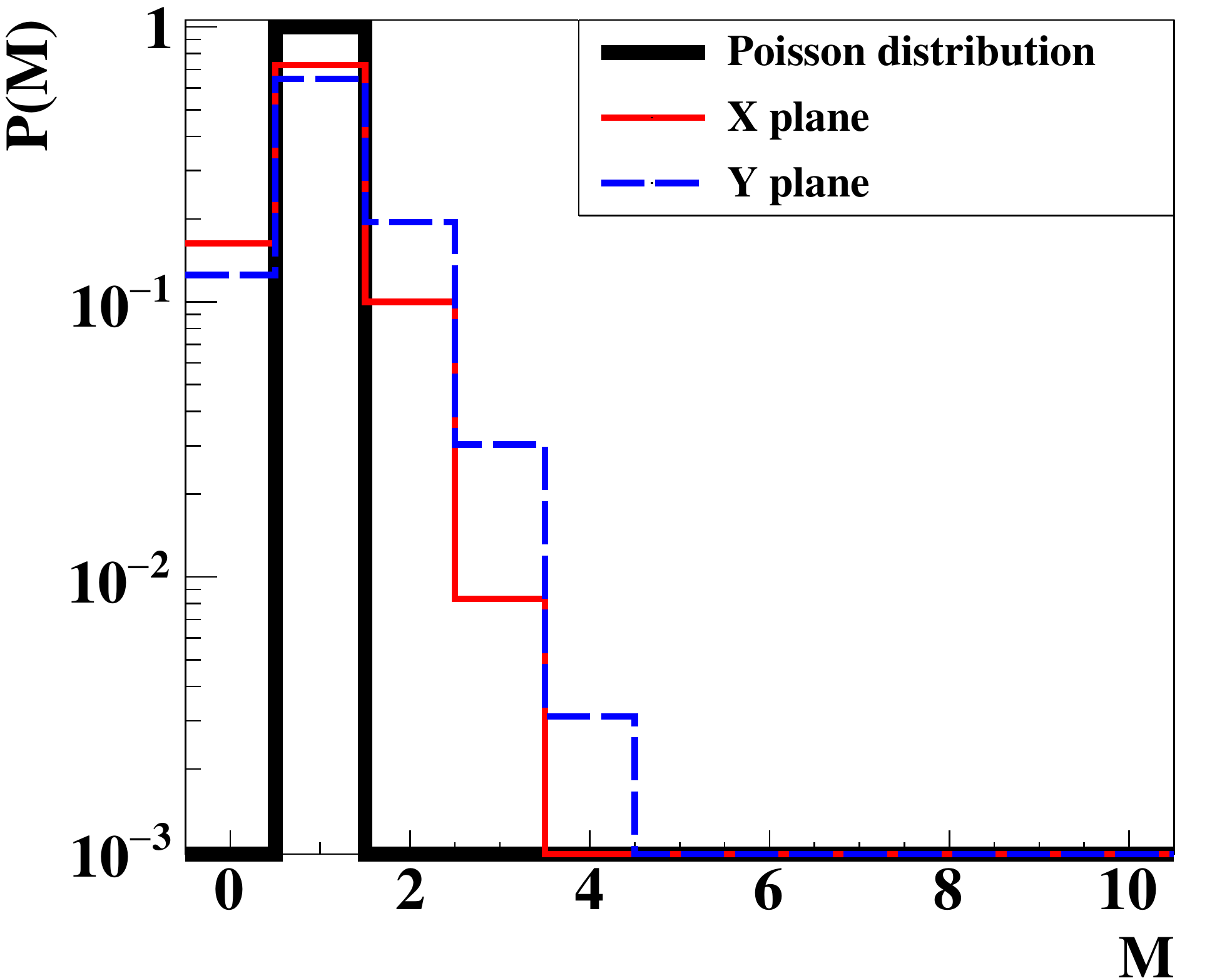} \caption{} \label{fig:Fibers_17kHz}
    \end{subfigure}
    \begin{subfigure}{0.32\textwidth} \centering \includegraphics[width=\textwidth]{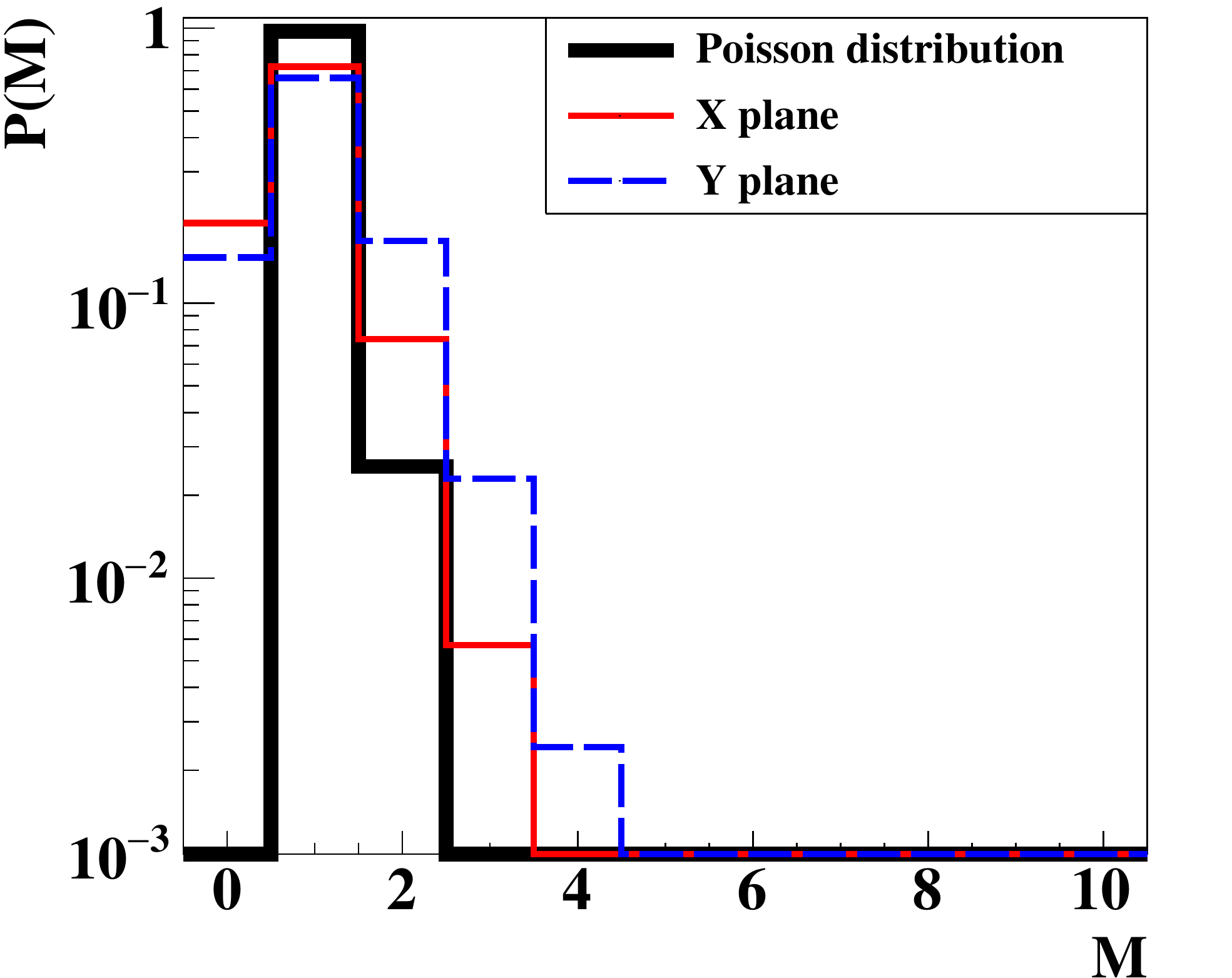} \caption{} \label{fig:Fibers_1MHz}
    \end{subfigure}
    \begin{subfigure}{0.32\textwidth} \centering \includegraphics[width=\textwidth]{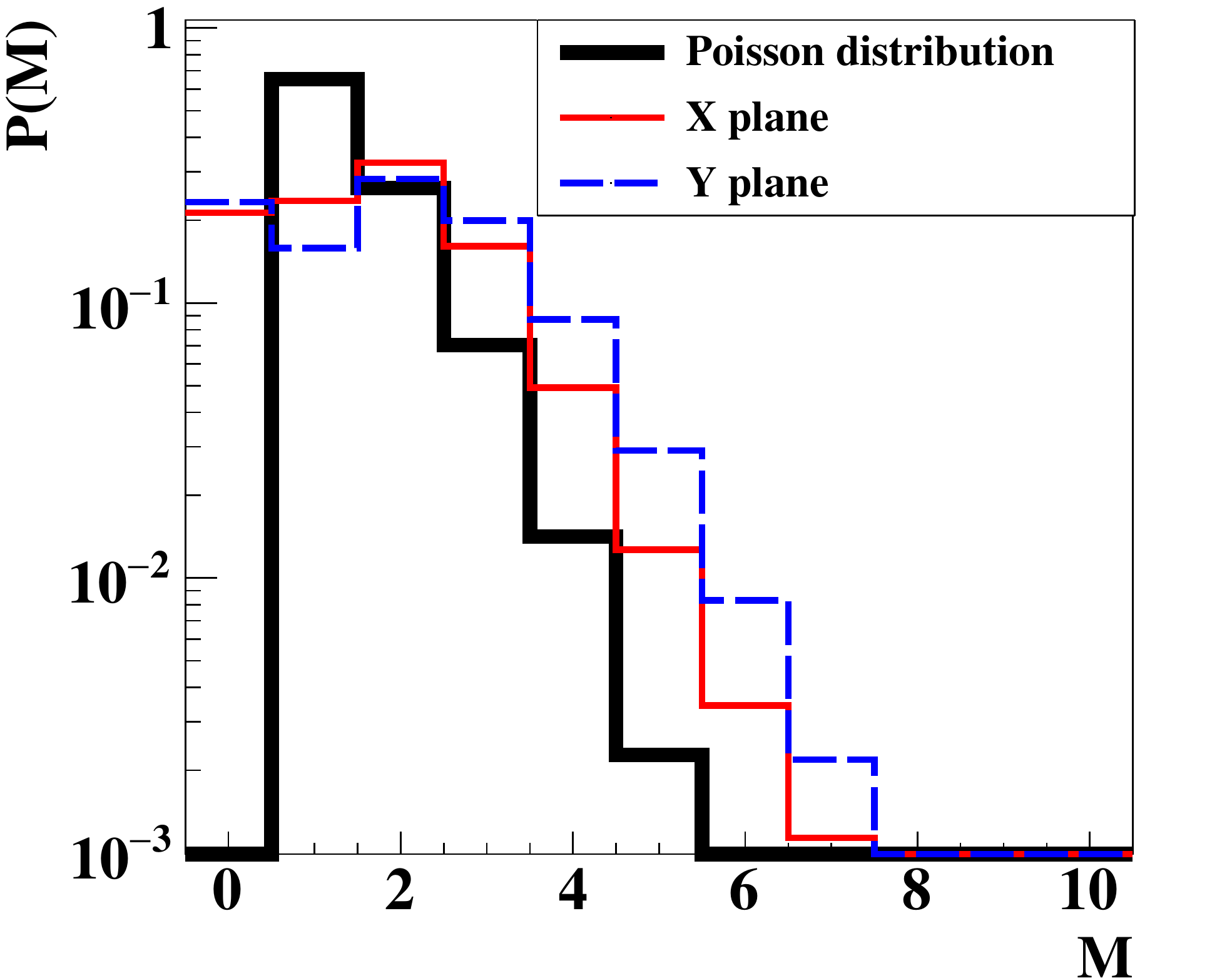} \caption{} \label{fig:Fibers_20MHz}
    \end{subfigure}
\caption{\small{\textit{Distribution of event multiplicity $P(M)$ for beam intensities of (a) 17~kHz, (b) $\sim$1.3 MHz and (c) $\sim$20 MHz.}} }
\label{fig:Multiplicity}
\end{figure}

Figure \ref{fig:Multiplicity} presents the distributions of multiplicity $P(M)$ of both planes for three beam intensities: 17~kHz, $\sim$1.3~MHz and $\sim$20~MHz. The mean number of protons per bunch is 6.8$\times$10$^{-4}$, 5.2$\times$10$^{-2}$ and 8.0$\times$10$^{-1}$ respectively. The Poisson distribution of the number of protons per bunch for bunches consisting of at least 1 proton is superimposed for each beam intensity since it corresponds to the expected multiplicity if we neglect the multiple proton arrivals in a single fiber. $P(M=0)=0$ since bunches without proton do not trigger the data acquisition. 

At low beams intensities (17~kHz and $\sim$1.3~MHz), we measured $P(M=1)$ values of 80\% and 70\% for X and Y planes respectively, to be compared to the expected value of 100\%. Experimental events with $M=0$ correspond to empty events including events with short pulse (ESP) (section~\ref{sec:hodo} and \ref{sec:Settings}). Events with $M\ge2$ could have been assigned to cross-talk between neighboring fibers. However since the distribution of distances between hit fibers is almost flat we concluded that these events are probably due to slight ground fluctuations in the ASICs leading to multiple hit channels (with “wrong hits”) while only one fiber has been crossed by an incident proton. 

At 20~MHz, the effect of ground fluctuations is amplified. Indeed, although the detection efficiency $P(M\ge1)$ remains almost constant, the experimental distribution of multiplicity significantly deviates from the expected distributions. Further improvements of the ASIC are therefore required to be compliant with beam intensities used in clinical centers.

\begin{figure}[htb]
\centering
    \begin{subfigure}{0.45\textwidth} \centering \includegraphics[width=\textwidth]{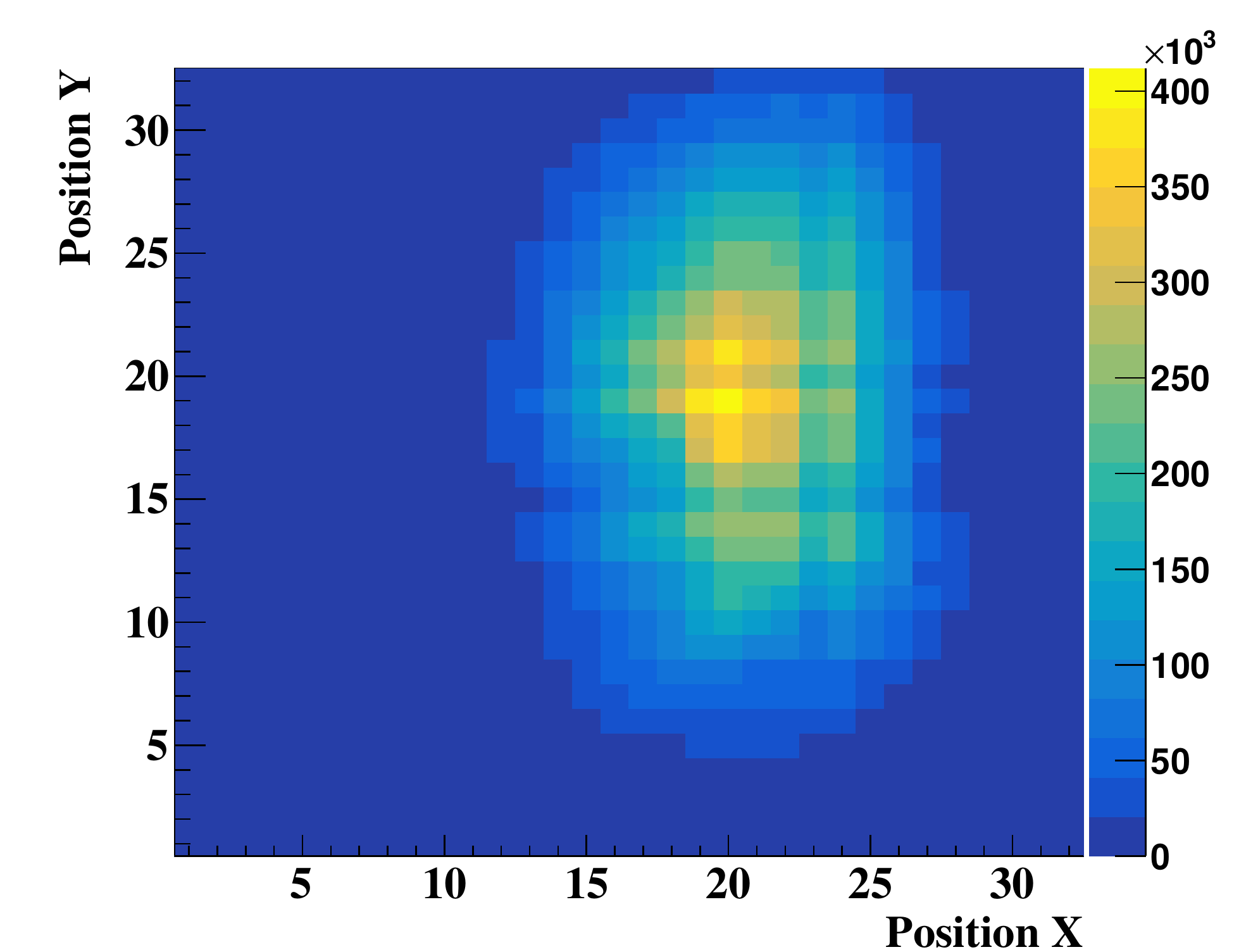} \caption{} \label{fig:2D_1MHz}
    \end{subfigure}
    ~
    \begin{subfigure}{0.45\textwidth} \centering \includegraphics[width=\textwidth]{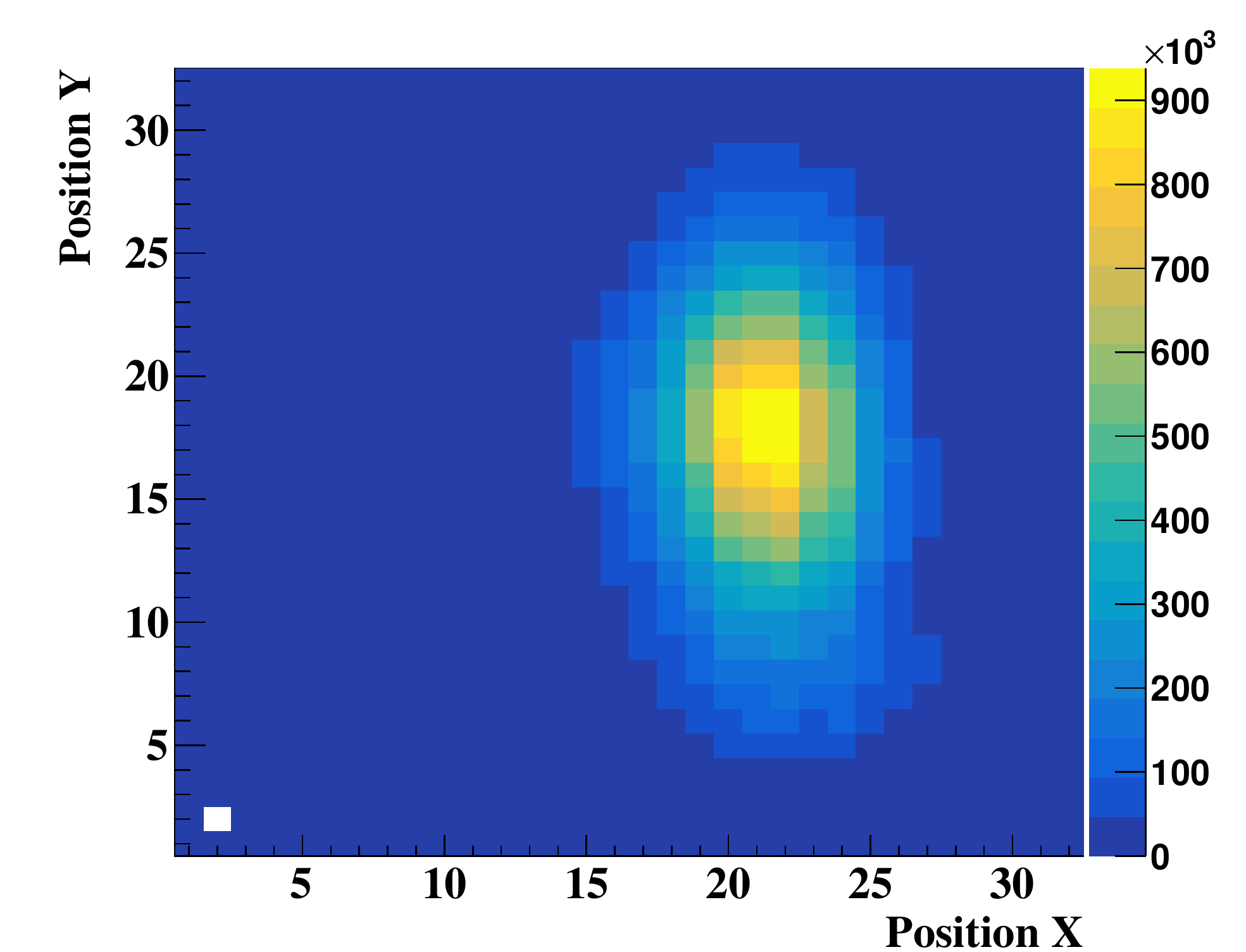} \caption{} \label{fig:2D_20MHz}
    \end{subfigure}
\caption{\small{\textit{2D map reconstructed from data collection of coincidence events (logical AND between X and Y planes) for beam intensities of (a) $\sim$1.3 MHz and (b) $\sim$20 MHz.}}}
\label{fig:2D_Maps}
\end{figure}

The monitoring software reconstructs two-dimensional maps for each acquisition. Figure~\ref{fig:2D_Maps} represents maps obtained for two acquisitions at $\sim$1.3~MHz and $\sim$20~MHz. Coincidence events (logical AND between both fiber planes) are reconstructed using the  average position in X and Y for events with $M>1$. Although the shape of the beam varies between the two intensities, the center of the beam is clearly defined. The shape modification is attributed to the change of the focusing lens of the beam when the intensity increases. 

\begin{figure}[htb]
\centering
    \begin{subfigure}{0.47\textwidth} \centering \includegraphics[width=\textwidth]{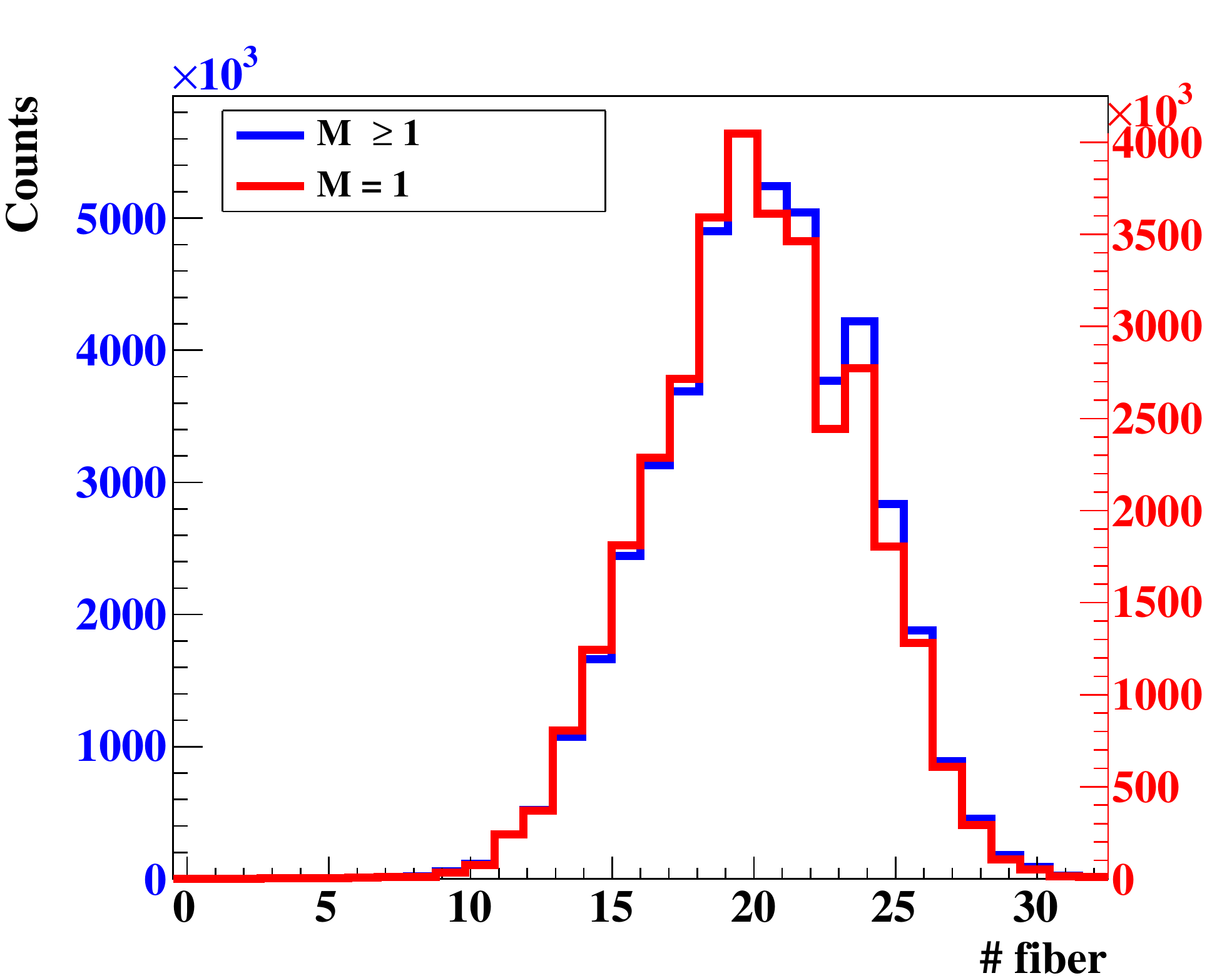} \caption{} \label{fig:Plane_X_1MHz}
    \end{subfigure}
    ~
    \begin{subfigure}{0.47\textwidth} \centering \includegraphics[width=\textwidth]{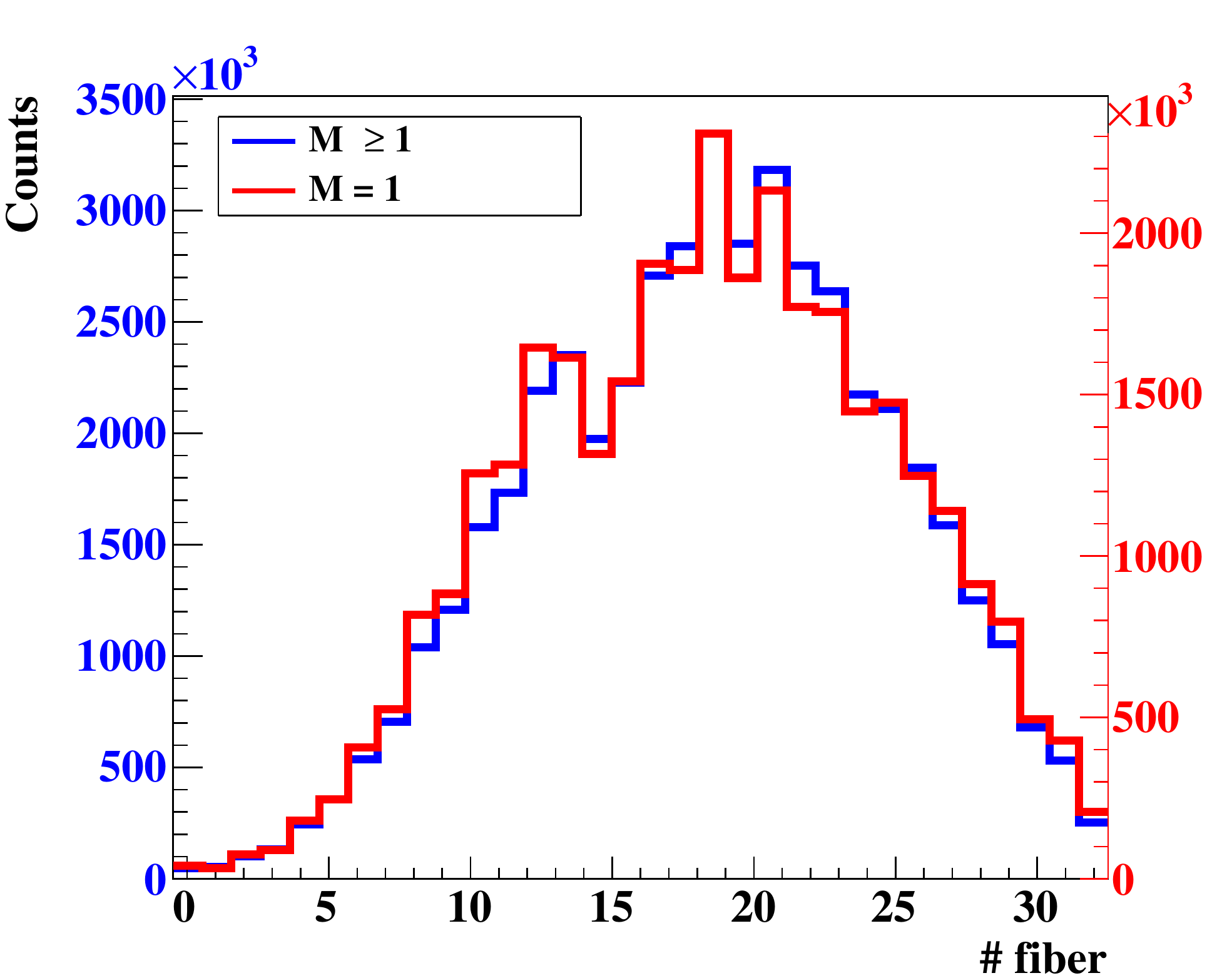} \caption{} \label{fig:Plane_Y_1MHz}
    \end{subfigure}
\caption{\small{\textit{1D profiles of (a) X and (b) Y fiber planes with multiplicities M=1 and M$\geq$1 for a beam intensity of $\sim${1.3}~MHz.}}}
\label{fig:1D_Profiles}
\end{figure}

Figure \ref{fig:1D_Profiles} represents the X and Y profiles obtained with two events selection methods for the irradiation with a beam intensity of $\sim$1.3~MHz. The blue curves correspond to the method used  for the 2D map reconstruction (average position of events with $M>1$) while the red curve has been obtained with the selection of $M=1$ events. The good overlap between the curves in both planes confirms that the method used for the 2D map reconstruction is adapted. However, for intensities larger than $\sim$20~MHz, the overlap of the two curves is lost due to the increase of \enquote{wrong hits} (not shown).

Considering that no structure is expected in the beam shape, the structures observed in the 1D profiles are attributed to slightly different efficiencies over the X and Y planes. The order of magnitude of these differences has been assessed focusing on the fibers obviously under-responding, namely fiber 23 in X plane and fibers 15, 16 and 20 in Y plane. Applying a linear interpolation over these fibers, we derived an estimate of the efficiency loss due to these fibers. Overall, the loss increases with the beam intensity from 0.75\% to 4\% and 2.7\% to 4\% in the X and Y planes, respectively.  

\subsubsection{Detection efficiency}

\begin{figure}[htb]
\centering
\includegraphics[width=0.7\textwidth]{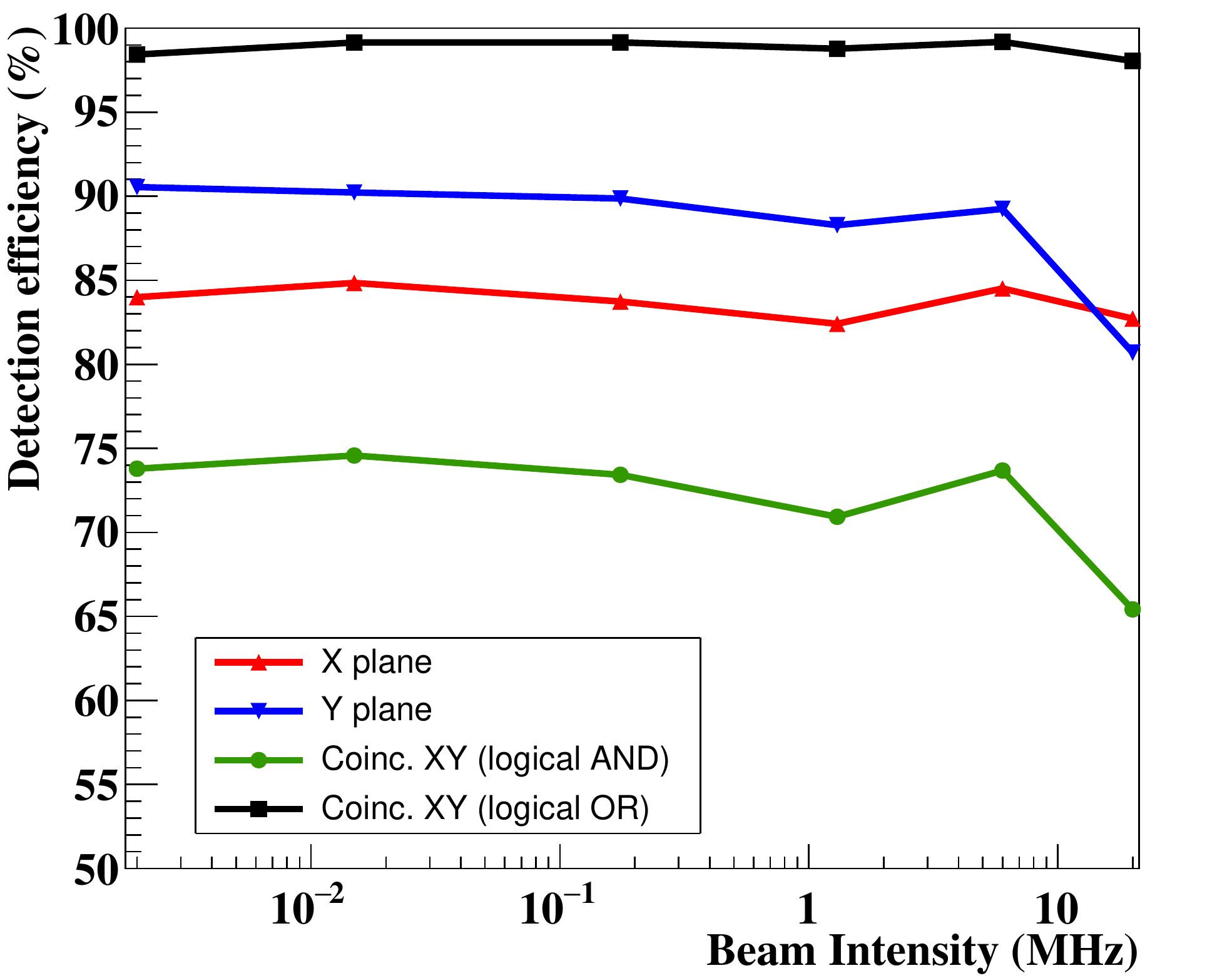}
\caption{\small{\textit{Hodoscope  detection  efficiency  as  a  function  of the proton beam intensity. A correction factor has been applied to take into account the loss of efficiency due to under-responding channels (see section~\ref{Profiles_And_Multiplies}).}}}
\label{fig:DE}
\end{figure}

Figure \ref{fig:DE} represents the detection efficiency of X and Y planes separately and with logical OR and AND conditions on the data of both the planes for various intensities from 2~kHz to $\sim$20~MHz. The data are corrected for under-responding channels (cf. \ref{Profiles_And_Multiplies}). The detection efficiencies in X and Y planes keep almost constant values around 84\% and 90\% respectively for intensities under $\sim$6~MHz. They result in a coincidence detection efficiency (logical AND between X and Y fibers) close to 74\% while the logical OR condition on both planes provides a detection efficiency close to 100\% in the same range of intensities. A significant fall-off of detection efficiency is observed at $\sim$20~MHz due to current ASIC limitations, especially ground fluctuations leading to data acquisition retriggering and consequently dead time.   

\subsubsection{Time resolution}

\begin{figure}[htb]
\centering
    \begin{subfigure}{0.3\textwidth} \centering \includegraphics[width=\textwidth]{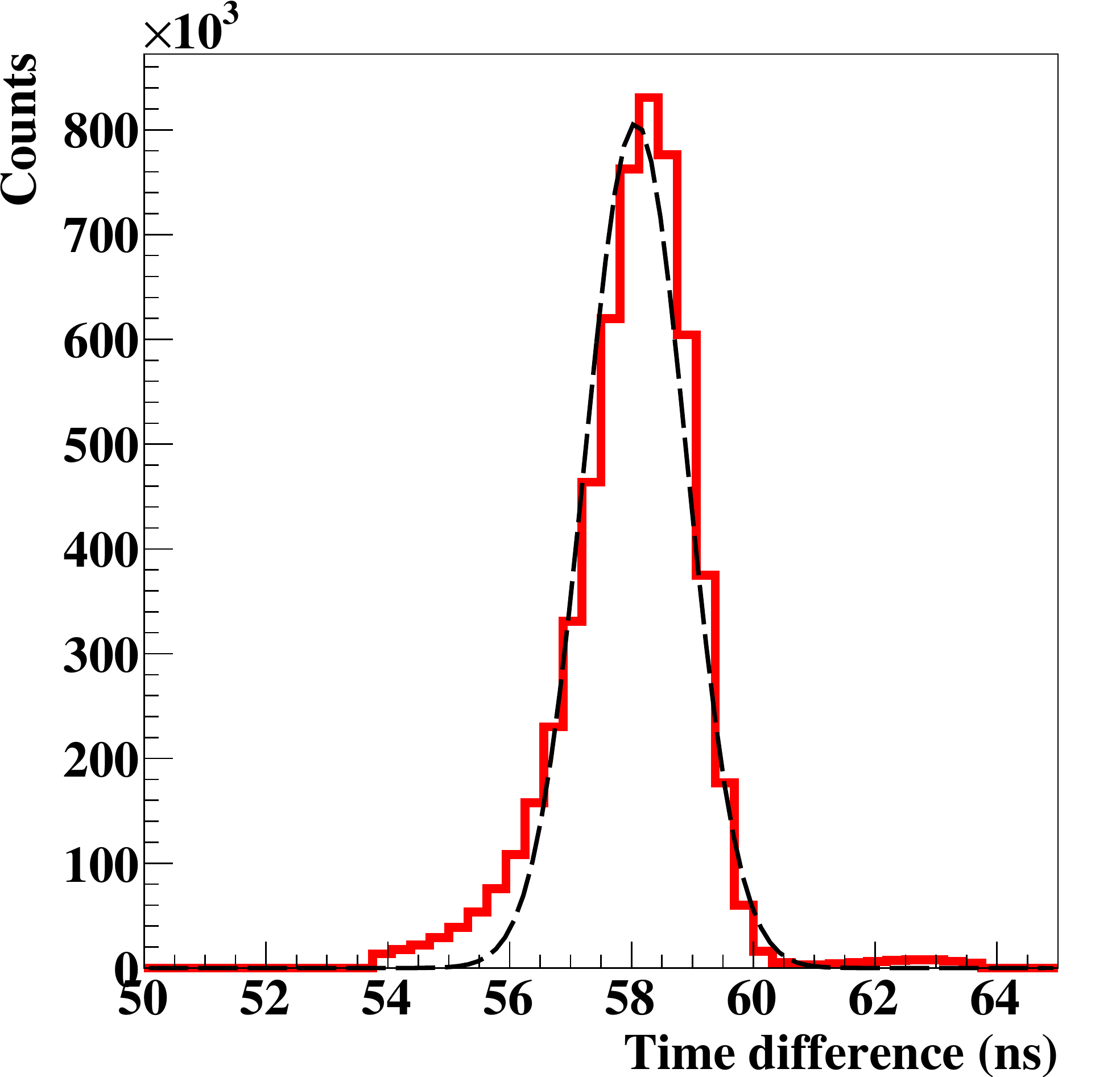} \caption{} \label{fig:Time_43kHza}
    \end{subfigure}
    ~
    \begin{subfigure}{0.3\textwidth} \centering \includegraphics[width=\textwidth]{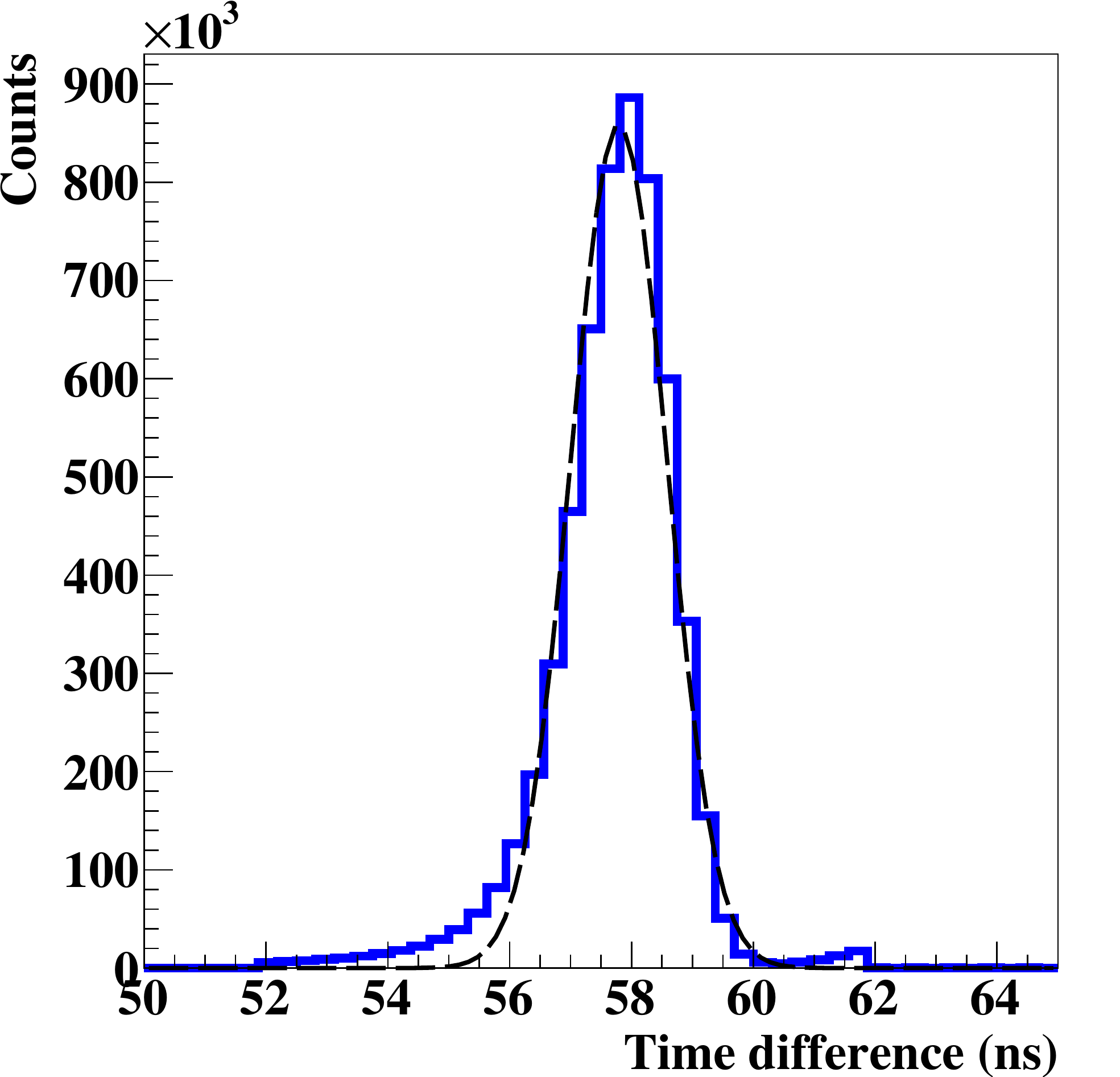} \caption{} \label{fig:Time_43kHzb}
    \end{subfigure}    
    ~
    \begin{subfigure}{0.3\textwidth} \centering \includegraphics[width=\textwidth]{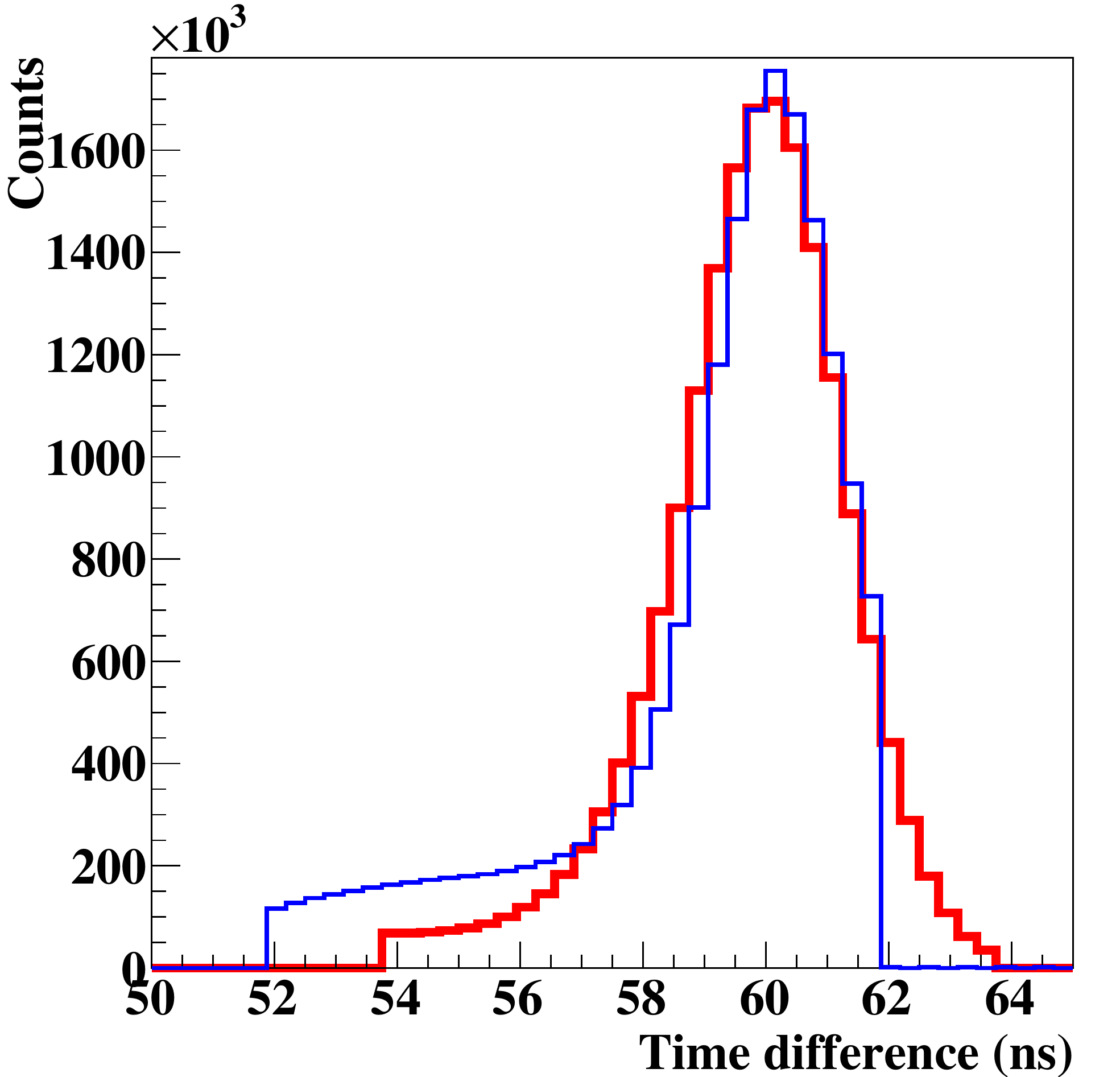} \caption{} \label{fig:Time_10MHz}
    \end{subfigure}
\caption{\small{\textit{Distribution of time difference between X and Y fibers (red and blue curves, respectively) and the trigger signal for beam intensities of 43~kHz ((a)--(b)) and $\sim$10~MHz (c).}}}
\label{fig:Time_coinc}
\end{figure}

Figure~\ref{fig:Time_coinc} shows the distributions of the time difference between the trigger signal and the first logic signal associated to each fiber plane (X and Y planes)  for beam intensities of 43~kHz (\ref{fig:Time_43kHza} and \ref{fig:Time_43kHzb}) and $\sim$10~MHz (\ref{fig:Time_10MHz}). The distributions obtained with a beam intensity of 43~kHz present well-defined peaks whose widths have been assessed employing a Gaussian fit. The measured full widths at half maximum (FWHM) are close to 1.8~ns which fulfills the specifications. The distributions at $\sim$10~MHz reveal an additional component before the main peak that is due to ASIC oscillations. Note that the time distributions are governed not only by the time resolution of the hodoscope but also by the ones of the PS detectors, which have not been measured.
\label{Time_resolution}

\section{Discussion}

The beam tagging hodoscope under development intends to provide incident ion detection with an efficiency larger than 90\% with a time resolution below 2~ns FWHM. The life duration of the scintillating fibers should be typically larger than 1000 patient treatments and the device should be capable to cope with 100~MHz counting rates expected with beam microstructures encountered in treatment centers. 

Regarding radiation hardness, the experiment performed at GANIL represents the worst case for radiation damage in hadrontherapy (carbon ions with less than 3~cm range in water). During a single irradiation experiment, less than 10\% efficiency decrease was observed with (3.6$\pm$0.7)$\times$10$^{12}$~ions$\cdot$cm$^{-2}$, which represents more than 1000 clinical irradiations for the central hodoscope area, without accounting for the longer time scale recovery. Comprehensive studies have been reported in \cite{Joram2015} and \cite{EkelhofPhD}. For instance Joram et al. mention a dose of 50~kGy beyond which LHCb scintillating fibre suffer from scintillation light yield degradation. Although our criterion to estimate fiber degradation is slightly different it is worth noting that the aforementioned fluence of 95~MeV/u carbon ions corresponds to a dose of $\sim150$~kGy which is of the same order of magnitude of the one reported by Joram et al.

In the meantime, significant progress has been made in the development of FE electronics and data acquisition system in order to fulfill the specifications in terms of detection efficiency, time resolution and counting rate capabilities. Moreover, specific configuration methods of the acquisition boards have been defined. Although the detection efficiency with coincidence between the X and Y planes is $\sim75$\%, it should be noticed that more than 90\% of the incident ions are detected by at least one of the fiber planes. Last but not least, the timing resolution is slightly better than the objective of 2~ns FHWM. 

However, the current limitations of the \enquote{HODOPIC} ASIC prevent us from reaching the counting rate of 100~MHz defined in the specifications. These limitations are probably due to ground oscillations in the ASICs that especially lead to fake channel hits. Such hits were shown in the experimental distributions of multiplicities compared to the Poisson distributions of the expected number of incident protons per bunch. They have also been observed in the distributions of time differences measured with a beam intensity of $\sim$10~MHz (figure \ref{fig:Time_10MHz}).

Although the time resolution of the CLaRyS hodoscope is slightly larger than the one of other prototypes developed worldwide, it is close to the expectations and further improvements can be foreseen with an upgrade of the ASIC. Moreover, the detection efficiency of a single fiber plane is comparable to the one of other prototypes for which the detection efficiency with a logical AND between two planes has not been reported yet.

\section{Conclusion}

The performances of the CLaRyS beam tagging hodoscope have been assessed in terms of detection efficiency, time resolution, event multiplicity and radiation hardness during several in-beam tests. A methodology for the configuration of the ASIC thresholds and channel gains has been tested and it allowed us to obtain a detection efficiency larger than 98\% with a logical OR and 72\% with a logical AND between X and Y fiber planes as well as a time resolution lower than 2~ns FWHM for intensities under 1~MHz. In conclusion, these performances are in accordance with the state of the art. Further improvements of the \enquote{HODOPIC} ASIC are required to reach the counting rate capability defined in the specifications (100~MHz).


\acknowledgments
This work was partially performed in the framework of Labex PRIMES (ANR-11-LABX-0063) and within the frame of the EU Horizon 2020 project RIA-ENSAR2/MediNet (654 002).  




\end{document}